\documentclass[10pt,A4paper]{article}
\usepackage{geometry}
\usepackage{amsfonts}
\usepackage{amsmath}
\usepackage{amssymb}
\usepackage{mathrsfs}
\usepackage{graphicx}
\usepackage{hyperref}
\usepackage[usenames,dvipsnames]{color}
\usepackage{cancel}
\usepackage{cleveref}
\usepackage{bbm}
\usepackage{bm}
\usepackage{mathtools}
\usepackage{physics}
\usepackage{tensor}
\usepackage{array}
\usepackage{float}
\usepackage{wrapfig}
\usepackage{soul}
\usepackage{xcolor}
\usepackage{cite}
\usepackage{url}

\hypersetup{
colorlinks=true,
linkcolor=blue,
citecolor=red,
urlcolor=cyan
}

\title{\vspace{-1cm} \bf Quantum Euler angles and\\agency-dependent spacetime}
\date{}
\author{{\large G. Amelino-Camelia,$^{(1)}$ V. D'Esposito,$^{(1)}$ G. Fabiano,$^{(1)}$}\\{\large D. Frattulillo,$^{(1)}$ P. A. H\"ohn,$^{(2)}$ F. Mercati$^{(3)}$\thanks{\href{mailto:flavio.mercati@gmail.com}{flavio.mercati@gmail.com}}
}
\vspace{12pt}
\\
\small $^{(1)}$Dipartimento di Fisica Ettore Pancini, Universit\`a di Napoli ``Federico II'';
\\
\small and INFN, Sezione di Napoli, Complesso Univ. Monte S. Angelo, I-80126 Napoli, Italy;
\\
\small $^{(2)}$Qubits and Spacetime Unit, Okinawa Institute of Science and Technology (\raisebox{-0.08em}{\includegraphics[height=0.8em]{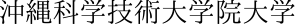}}),
\\
\small  1919-1 Tancha, Onna-son, Kunigami-gun, Okinawa, Japan 904-0495;
\\
\small $^{(3)}$Departamento de F\'isica, Universidad de Burgos, 09001 Burgos, Spain.
}

\begin{document}

\maketitle
\begin{abstract}
Quantum gravity is expected to introduce quantum aspects into the description of reference frames. Here we begin exploring how quantum gravity induced deformations of classical symmetries could modify the transformation laws among reference frames in an effective regime. We invoke the quantum group $SU_q(2)$ as a description of deformed spatial rotations and interpret states of a representation of  its algebra as describing the relative orientation between two  reference frames.
This leads to a quantization of one of the Euler angles and to an aspect of \emph{agency dependence}: space is reconstructed as a collection of fuzzy points, exclusive to each agent, which depends on their choice of reference frame. Each agent can choose only one direction in which points can be sharp, while points in all other directions become fuzzy in a way that depends on this choice. Two agents making different choices will thus observe the same points with different degrees of fuzziness.
\end{abstract}

\maketitle

\vspace{1cm}

\section{Agency-dependent spacetime and spacetime-dependent agency}

{Observers typically play a somewhat ``external'' role in both general relativity (GR) and quantum theory. In the former, they are  assumed to exert a negligible backreaction on spacetime, which they can thus probe without influencing it. In quantum theory, it is the Heisenberg cut \cite{heisenberg1949physical, wiseman2009quantum} that separates them from the observed system.} The observers' knowledge and choices are described classically, however, there are mutually-incompatible measurements that they can choose to perform (complementarity). Quantum mechanics introduces therefore an important novelty: \emph{the operationally-meaningful properties of the observed systems depend on the choices made by the observer.} Through the impact of these choices, observers are ``agents" for what concerns measurements in quantum mechanics.

{When quantum theory and GR are combined in a quantum theory of gravity, one ends up considering spacetime as a quantum object.} We may then contemplate the possibility that the agency-dependence of quantum mechanics might  challenge the notion of an objective spacetime that all observers agree upon: how spacetime reveals itself to observers may depend on some of their choices.
Conversely, the quantum properties of spacetime will likely affect the spectrum of possible operations available to an agent. These observations  suggest a picture in which spacetime and the agency of observers affect each other inextricably, so much so that the ``externality'' idealization, a good working hypothesis in GR  and quantum theory, will have to be abandoned in favor of a notion of ``internal'' observers in quantum gravity \cite{Hohn:2017cpr}.

In the present paper, we take a step toward this largely unexplored aspect of quantum gravity. We cannot expect to begin with a complete model of an observer/agent in a quantum spacetime. Therefore, we will be concerned with a particular aspect of this problem: the relative state of the reference frames of two observers, Alice and Bob. In a simplified model, we assume that the two observers are exclusively interested in finding out the relative spatial orientations of their respective laboratories. In standard quantum theory, the rotation matrix relating the two reference frames can be determined by exchanging qubits and classical information, as discussed in \cref{Protocol sec}. With an arbitrarily large number of exchanges, the two observers can determine the matrix elements with arbitrary precision. However, in the case in which Alice and Bob live in a quantum spacetime, reaching an arbitrary precision may be forbidden. 

\subsection*{Quantum group toy model}
We will focus on a toy model of quantum gravity in the form of a nonclassical space: this will be the arena in which observers exercise their agency and will attempt to determine the structure of the spacetime they are in. A much-studied candidate for a theory of quantum spacetime is \emph{noncommutative geometry} \cite{Szabo:2009tn}. The idea is that, by replacing the (commutative) algebra of functions on a manifold with a noncommutative algebra, we end up describing a manifold whose points are ``fuzzy", due to the Heisenberg uncertainty principle. In this sense, standard quantum theory can be seen as the noncommutative space one obtains by making the algebra of phase-space functions noncommutative. Recent developments focused on deformed symmetries and \emph{quantum groups}, in which the algebra of functions on a Lie group is noncommutative~\cite{majid_2002}. As Lie groups describe the symmetries of classical spaces, quantum groups describe the symmetries of noncommutative geometries; as such they have appeared in several quantum gravity approaches \cite{Amelino-Camelia:2002cqb, Smolin:2002sz, Freidel:2005me, Amelino-Camelia:2007une, Girelli:2007xn, Girelli:2009yz, Bianchi:2011uq, Dupuis:2013lka, Dupuis:2020ndx, Ballesteros:2021bhh, Calmet:2021sws, Girelli:2022foc, Bonzom:2022bpv, Arzano:2022nlo}. In this regard, since we are only interested in the relative orientation between reference frames, we can model our nonclassical space simply by deforming the rotation group $SO(3)$, or rather its double cover $SU(2)$ which describes the rotations of qubits.

To do so, we  will be concerned with one of the simplest examples of quantum groups: $SU_q(2)$~\cite{Woronowicz87}, which is the (only) quantum-group  deformation of the  $SU(2)$ group. The lowercase $q$ identifies a dimensionless deformation parameter, such that the case $q=1$ reproduces the undeformed $SU(2)$ group. In loop quantum gravity models, $q$ is a function of the dimensionless ratio between the Planck length and the Hubble length scale associated to the cosmological constant \cite{Smolin:2002sz}. It has been argued that this reflects a minimal possible resolution in angular measurements~\cite{Bianchi:2011uq}.  One could also imagine a physical scenario in which the dimensionless $q$ could acquire a dependence on the characteristic energy scale $E$ of the problem via, for example, the dimensionless ratio $E/E_p$, with $E_p$ the Planck energy. This feature could arise in field theories in which $q$ is a parameter appearing in the lagrangian, and thus would run with the energy as other parameters do. To the best of our knowledge, no concrete example with such features has been constructed, so we do not speculate upon this possibility further.

The mathematical details of $SU_q(2)$ have been developed extensively \cite{Woronowicz87, Vaksman1988, Masuda1991, Podles1995}. We want to further explore the physical features of this group to arrive at a qualitative prediction for the novel ``quantum geometrical'' effects that observers would see if, in an effective regime of quantum gravity, rotations of reference frames were described by $SU_q(2)$ transformations.

\section{Quantum rotation matrices in $SU_q(2)$}
\label{SUq2Sec}

\subsection{$SU_q(2)$ algebra and homomorphism with $SO_q(3)$}\label{suq2}

The quantum group $SU_q(2)$ is defined by considering the algebra of complex functions on $SU(2)$, denoted by $C(SU(2))$ and deforming it in a noncommutative way. The generators $\{a,c\}$ are regarded as functions on the group, satisfying non-trivial commutation relations
\begin{equation}
\begin{aligned}
& ac=qca \qquad a c^*=q c^*a \qquad cc^*=c^*c \\
& c^*c+a^*a = \mathbbm{1} \qquad aa^*-a^*a=(1-q^2)c^*c\,.
\end{aligned}
\label{deformed group algebra}
\end{equation}
Here, $\mathbbm{1}$ refers to the identity element of the algebra and $q$ is the deformation parameter assumed to be close to $1$ and in particular $q \lesssim 1$. Indeed, in the $q\rightarrow 1$ limit, we obtain the commutative limit and recover the classical description of $SU(2)$. In passing, we mention that it suffices to consider $q \lesssim 1$ since, if $q>1$, the mapping $a \mapsto a^*$, $c\mapsto qc^*$ sends the $SU_q(2)$ algebra to the $SU_{q^{-1}}(2)$ one.

To establish a first link with the classical picture, we present the generalization of the spin-$1/2$ representation \eqref{su2classica}, given by
\begin{equation}
    U_q=\begin{pmatrix}\label{su2quantum}
    a & -qc^*\\
    c & a^*
\end{pmatrix} \qquad a,c\in C(SU_q(2)) \quad q\in(0,1)\,.
\end{equation}
We will now construct one of the key ingredients of the analysis which is a $q$-analogue of the $SU(2)$ to $SO(3)$ homomorphism. In order to do so, we promote the classical homomorphism  $\tensor{R}{_{ij}} =  \frac{1}{2}\Tr\qty{\sigma_j\,U^\dagger \, \sigma_i\, U}$, where $\sigma_i$ are the Pauli matrices, to its quantum version by replacing $U$ with $U_q$
\begin{equation}
\label{trace}
  \tensor{\qty(R_q)}{_{ij}} =  \frac{1}{2}\Tr\qty{\sigma_j\,U_q^\dagger \, \sigma_i\, U_q}\,.
\end{equation}
By computing these elements explicitly, we obtain

\begin{equation}
    R_q = \begin{pmatrix}
    \frac{1}{2}(a^2-qc^2+(a^*)^2-q(c^*)^2) & \frac{i}{2}(-a^2+qc^2+(a^*)^2-q(c^*)^2) & \frac{1}{2}(1+q^2)(a^*c+c^*a) \\
    \frac{i}{2}(a^2+qc^2-(a^*)^2-q(c^*)^2) & \frac{1}{2}(a^2+qc^2+(a^*)^2+q(c^*)^2) & -\frac{i}{2}(1+q^2)(a^*c-c^*a) \\
    -(ac+c^*a^*) & i(ac-c^*a^*) & 1-(1+q^2)cc^*
    \end{pmatrix}
    \label{matrice rotazione quantum 1}
\end{equation}
and one can check that in the commutative limit $q\rightarrow 1$ this coincides with the matrix obtained via the classical homomorphism.
This matrix provides a (co)-representation of $SO_q(3)$  related to the (co)-representation found in \cite{Podles1995} by a similarity transformation. We note that in \eqref{trace} the cyclic property of the trace operation does not hold because of the noncommutative nature of $a$ and $c$. This gives rise to a ``quantization ambiguity" which however does not affect our phenomenological results. The relevant technical issues are addressed in \cref{Quantization ambiguity sec}.

The classical analog of \eqref{matrice rotazione quantum 1} describes all possible elements of $SO(3)$ when varying the complex numbers $a$ and $c$ with continuity. Each of these classical matrices describes a possible relative orientation between observers A and B. In the quantum case \eqref{matrice rotazione quantum 1} has no physical meaning when taken alone, but only when paired with a certain state $\ket{\psi}\in\mathcal{H}$, where $\mathcal{H}$ is the Hilbert space on which $a$ and $c$ act. In the classical case, $a$ and $c$ codify information about the alignment of two reference frames, via their angular parametrization (see below and \cref{Quantization ambiguity sec}). For this reason, when being promoted to operators, we interpret the states on which they act as the ones codifying the relative orientation between the reference frames of observers $A$ and $B$. The physical information about this orientation can be extracted by computing the expectation values $\expval{\tensor{(R_q)}{_{ij}}}{\psi}$ with their relative uncertainties stemming from the noncommutative nature of these matrix elements, given by

\begin{equation}
\label{incertezze}
\Delta_{ij} = \sqrt{ \expval{(R_q)_{ij}^2}{\psi} -\expval{(R_q)_{ij}}{\psi}^2}\,.
\end{equation}
In this framework, the $\Delta_{ij}$ do not vanish simultaneously, in general, for a given state $\ket{\psi}$ introducing a ``fuzziness" in the alignment procedure. When this occurs, the latter is affected by an intrinsic uncertainty which cannot be eliminated: A and B are not able to sharply align their reference frames anymore. In light of these arguments, it is crucial to identify $\mathcal{H}$ and study the representations of operators $a$ and $c$ on it.

\subsection{$SU_q(2)$ representations and quantum Euler angles}\label{reps}

The representations of algebra \eqref{deformed group algebra} have been thoroughly studied in \cite{Podles1995}.
The Hilbert space containing the two unique irreducible representations of the $SU_q(2)$ algebra, $q\in(0,1)$, is $\mathcal{H} = \mathcal{H}_{\pi} \oplus \mathcal{H}_\rho$ where $\mathcal{H}_{\pi} = \ell^2\otimes L^2(S^1)\otimes L^2(S^1)$ and $\mathcal{H}_{\rho} = L^2(S^1)$. If $\chi, \phi \in [0,2\pi[$ are coordinates on $S^1$ and $\ket{n}$ is the canonical basis of $\ell^2$, the algebra of functions on $SU_q(2)$ is represented as
\begin{equation}
    \rho(a) \ket{\chi}=e^{i\chi}\ket{\chi}\qquad \rho(a^*)\ket{\chi}=e^{-i\chi}\ket{\chi} \qquad \rho(c)\ket{\chi}=\rho(c^*)\ket{\chi}=0
    \label{rappresentazione rho e rhostar}
\end{equation}

\begin{eqnarray}
    \pi(a)\ket{n, \phi, \chi}&=&e^{i\chi}\sqrt{1-q^{2n}}\ket{n-1, \phi, \chi} \qquad \pi(c)\ket{n, \phi, \chi}=e^{i\phi}q^{n}\ket{n, \phi, \chi}
    \nonumber\\ \label{rappresentazione pi}\\
    \pi(a^*)\ket{n, \phi, \chi}&=&e^{-i\chi}\sqrt{1-q^{2n+2}}\ket{n+1, \phi, \chi} \qquad \pi(c^*)\ket{n, \phi, \chi}=e^{-i\phi}q^{n}\ket{n, \phi, \chi}\,.\nonumber
\end{eqnarray}
It is not coincidental that quantum number $\chi$ is common in both representations. We show in \cref{limit sec} that representation $\rho$ can be obtained as a limit of representation $\pi$, in agreement with the fact that in the classical case we only need three real parameters to specify rotations.

\begin{figure}[h!]
    \centering
    \includegraphics[scale=0.4]{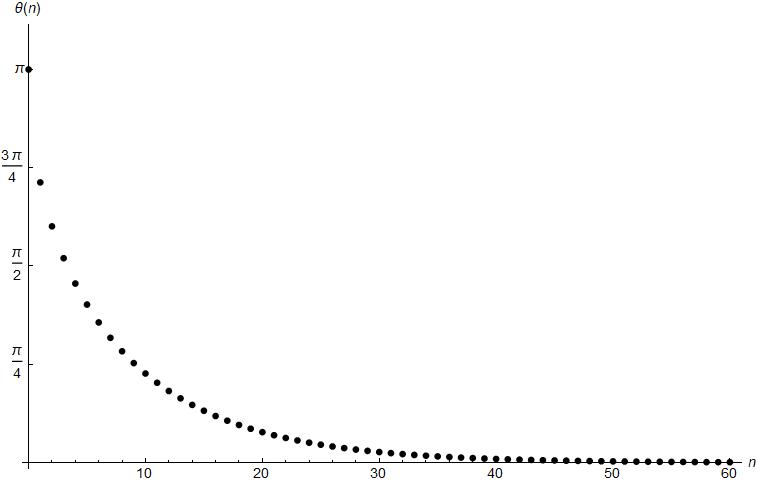}
    \caption{Discretized angles \eqref{angoli quantum} computed with $q=0.9$. $\theta(n)$ is decreasing with $n$ starting from $\pi$ and approaching $0$ for large values of $n$. Interestingly, the step between consecutive angles decreases as $n$ gets very larger, and for very large $n$
 we can approximate the angular distribution as being continuous. This can easily be verified analytically by computing $\Delta \theta(n)=\theta(n+1)-\theta(n)$ and taking the limit for $n\rightarrow \infty$. Another feature that gives robustness to our proposal is that the angular steps get smaller as $q$ is closer to $1$.}
    \label{angoli}
\end{figure}

Let us give physical meaning to the quantum numbers appearing in the representations, making a comparison between these relations and the classical parametrization of $a$ and $c$ for $U \in SU(2)$

\begin{equation}
     a = e^{i\eta}\cos{\frac{\theta}{2}} \qquad c = e^{i\delta} \sin{\frac{\theta}{2}}  \,,
    \label{rappresentazione classica a e c}
\end{equation}
where $\eta$, $\delta$ and $\theta$ are a linear redefinition of Euler angles, as discussed in \cref{Quantization ambiguity sec}. We note that $\rho(a)$ and $\rho(c)$ act diagonally, therefore we can make the identifications $\chi\equiv\eta$ and $\theta\equiv0$.
For what concerns representation $\pi$, $c$ acts diagonally, while $a$ and $a^*$ act as ladder operators. 
For operator $a$, we added a phase, that is implicitly set to $0$ in the literature \cite{Woronowicz87}, which does not affect the commutation relations \eqref{deformed group algebra} but is needed in order to  have a comparison of \eqref{rappresentazione pi} with \eqref{rappresentazione classica a e c}.
From this, we identify $\chi\equiv\eta$, $\phi\equiv\delta$ and, exploiting the fact that $c$ acts diagonally, we are led to the significant result
\begin{equation}\label{angoli quantum}
    q^n=\sin(\frac{\theta(n)}{2})\iff \theta(n)=2\arcsin(q^n)\,.
\end{equation}
\emph{The Euler angle $\theta$ becomes quantized}, a feature captured in \cref{angoli}. The quantization of $\theta$ can also be inferred using $\pi(a)$ (see \cref{Quantization ambiguity sec}). 
\subsection{Semi-classical rotations}\label{semiclassical rotations}

To gain further intuition from the classical picture, where the rotation axis and the angle by which the rotation is performed are specified by three Euler angles, we henceforth focus on``semi-classical'' rotations, specified by the three quantum numbers $\theta(n),\phi,\chi$, that describe small deformations of classical rotations defined by these angles. More precisely, the states $\ket{\psi(\theta,\phi,\chi)}$ yielding such deformed rotations should satisfy

\begin{equation}
\left\{\begin{array}{ll}
\expval{\qty(R_q)_{ij}}{\psi\qty(\theta,\phi,\chi)}= R_{ij}\qty(\theta,\phi,\chi)+O(1-q) \\ \\ \Delta_{ij}^2= O(1-q)
\end{array} \right. \qquad \forall \; i,j\,\,,
\label{criterio rotazione classica}
\end{equation}
where $R_{ij}\in\rm{SO}(3)$ and where $\theta$ is one of the allowed values in \eqref{angoli quantum}. This prescription constrains the quantum rotations we can construct to the ones in which the $\theta$ Euler angle can only take the allowed values in \eqref{angoli quantum}, for a fixed $q$.
In order to clarify the meaning of this prescription, we provide some examples. 

A first class of states that trivially satisfies requirement \eqref{criterio rotazione classica} is the set of eigenstates $\ket{\chi}\in\mathcal{H}_\rho$. It is easy to see that the expectation value of the quantum rotation matrix \eqref{matrice rotazione quantum 1} in such states describes a rotation of angle $2\chi$ around the $z$-axis, namely:
\begin{equation}\label{rotation z axis}
    \expval{R_q}{\chi}=\begin{pmatrix}
    \cos(2\chi) & -\sin(2\chi) & 0 \\
   \sin(2\chi) & \cos(2\chi) & 0 \\
  0 & 0 & 1
    \end{pmatrix}
\end{equation}
with all the $\Delta_{ij}$ being zero. Thus, rotations around the z-axis are classical/continuous:
two observers, A and B, whose relative orientation is described by $\ket{\chi}\in\mathcal{H}_\rho$, can align themselves sharply.

The relative simplicity of rotations around the z-axis is not representative of the richness of structure of other rotations, and this is mainly due to the fact that  for generic basis states of the form $\ket{n,\phi,\chi}$ 
the first condition in \eqref{criterio rotazione classica}
is not satisfied, forcing us to consider superpositions of such states. Since $\phi$ and $\chi$ are identified with their classical counterparts, we are led to consider superpositions 
\begin{equation}
\label{superposition}
    \ket{\psi}=\sum_{n=0}^{\infty}c_n\ket{n,\phi,\chi} \;\;\,\, , \;\;\,\, \sum_{n=0}^\infty \abs{c_n}^2=1 \; \;.
\end{equation}
Among the states of this form that satisfy  \eqref{criterio rotazione classica}, we need to find those for which the uncertainties are kept under control. The best way to do this is to demand that the coefficients $\{c_n\}$ minimize the functional 

\begin{equation}
\label{functional}
    S\Big[\{c_n\}_{n=0}^\infty,\mu\Big]=\sum_{i,j}\Delta_{ij}^2 - \mu \big(\braket{\psi}-1\big)\,,
\end{equation}
where $\mu$ is a Lagrange multiplier enforcing normalization of $\ket{\psi}$.
In general, solving the minimization problem \eqref{functional} is a daunting task computationally. Therefore, in \cref{Numerical construction sec} we invoke physical intuition to construct these states. Fixing a value for $q$, we build superpositions of states $\ket{n,\phi,\chi}$ centered around a certain $\Bar{n} \in \mathbb{N}$. The expectation values and uncertainties relative to such states will reproduce deformations of classical rotation matrices specified by angles $\qty(\theta(\Bar{n}),\phi,\chi)$, in the sense of \eqref{criterio rotazione classica}. 

It is worth noticing that there is a special class of such 
semi-classical rotations for which some simplifications arise. This is the case of rotations with $\theta = \pi$ around axes of the $x-y$ plane.
The classical theory predicts that $\chi=0$ and a generic $\phi$ select a direction in the $x-y$ plane ($\phi=\pi/2$ selects a rotation around the $x$-axis, while $\phi=0$ selects one around the $y$-axis) around which we rotate of an angle $\theta$. 
As emphasized before, since  $\phi,\chi$ behave classically, the quantum numbers associated to these angles specify the axis of rotation also in the quantum case. 
For the case $\theta(0)=\pi$, building a superposition of basis states is not necessary to satisfy \eqref{criterio rotazione classica} and the basis state  $\ket{0,\phi,0}$ is sufficient to describe the corresponding semi-classical rotation around an axis in the $x-y$ plane, specified by $\phi$.
The expectation value of $R_q$ on such state is 
\begin{equation}\label{axisplane}
    \expval{R_q}{0,\phi,0}=\begin{pmatrix}
    -q\cos(2\phi) & -q\sin(2\phi) & 0 \\
   -q\sin(2\phi) & q\cos(2\phi) & 0 \\
  0 & 0 & -1
    \end{pmatrix}
\end{equation}
while the $\Delta_{ij}$ are non-zero and do not depend on the specific value of $\phi$
\begin{equation}\label{varxyplane}
    \Delta R_q=\begin{pmatrix}
    \frac{1}{2}\sqrt{(1-q^2)(1-q^4)} & \frac{1}{2}\sqrt{(1-q^2)(1-q^4)} & \frac{(1+q^2)}{2}\sqrt{(1-q^2)} \\
    \frac{1}{2}\sqrt{(1-q^2)(1-q^4)} & \frac{1}{2}\sqrt{(1-q^2)(1-q^4)} & \frac{(1+q^2)}{2}\sqrt{(1-q^2)} \\
        \sqrt{q^2-q^4} & \sqrt{q^2-q^4} & 0
    \end{pmatrix}
\end{equation}
From the above matrices, it can be verified that the state $\ket{0,\phi,0}$ satisfy \eqref{criterio rotazione classica}. 

For a  generic axis of rotation, the quantum Euler angle $\theta$ will play a non-trivial role both in the determination of the axis itself and in the determination of the rotation angle.

\section{Same stars, different skies}
Perhaps the most striking implication of our results is that they provide a possible scenario for a novel type of relationship between observers and the spacetime they observe, such that the choices made by the observer in setting up her reference frame affect the spacetime she observes.

We shall argue this by first advocating Einstein's operational notion of spacetime, whose points have physical meaning only in as much as they label an event there occurring. We use as reference example a network of sources emitting photons (``stars"): each photon emission is a physical point of spacetime. These points of spacetime will be labeled by measured coordinates and uncertainties on those coordinates.
The key observation here is that, in our model, these uncertainties are not intrinsic properties of the spacetime points but rather depend on the choices made by the observer. These arguments give rise to a framework in which \emph{different skies may originate from the observation of the same stars}.

\subsection{Agency-dependent space}
We will apply our formalism to establish the connection between the $q$-deformation of the $SU(2)$ group and the agency-dependence of space. Before showing this with a concrete numerical example, we will discuss why this property of space is to be expected with a thought experiment.

Consider two observers, Alice and Bob, each equipped with their own set of telescopes, who want to map the starry sky. Each of them chooses their reference frame (in particular their $z$-axis) to assign coordinates to the stars they observe. We assume that their origins coincide but their $z$-axes do not. Without loss of generality we can focus on the $(y,z)$-plane so that, for each observer, any telescope is mapped onto another by a rotation about the $x$-axis. 
In the classical case, Alice and Bob can compare their experimental results by rotating their data with a classical rotation matrix. Formally, there exists an element of the $SO(3)$ group that sharply describes the relative orientation between Alice and Bob. 

\begin{figure}
    \centering
    \includegraphics[width=1\textwidth]{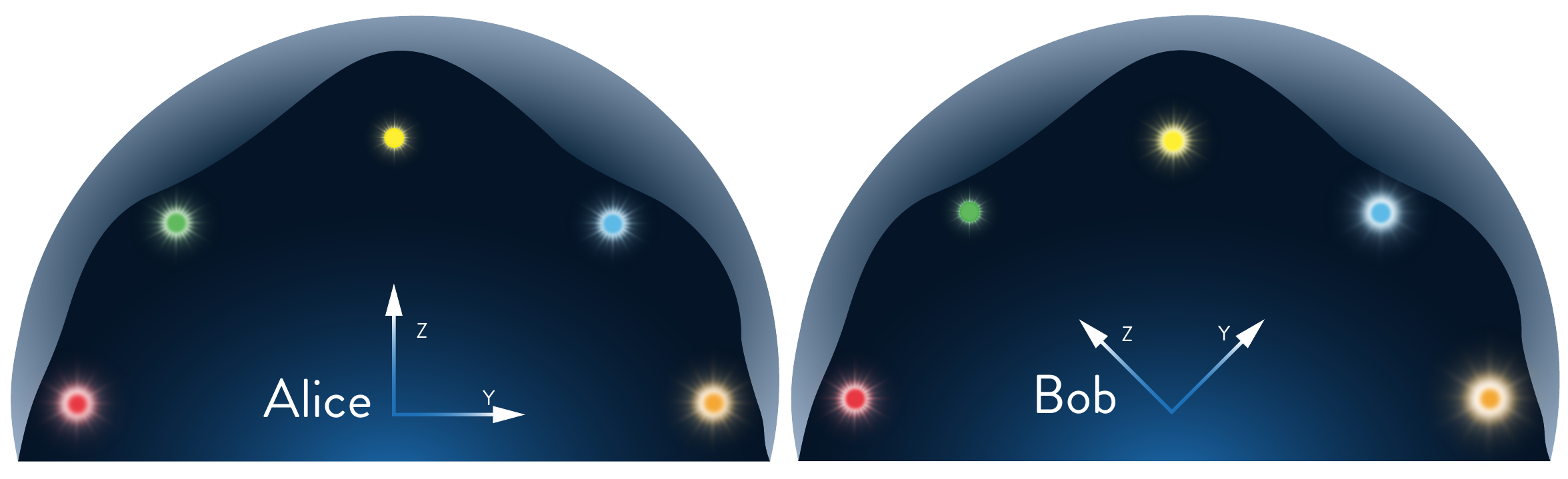}
    \caption{A semi-quantitative description of the starry skies observed by Alice and Bob. Alice's $z$-axis is aligned with the yellow star while Bob's $z$-axis is aligned with the green star, so that the yellow star is sharp for Alice while the green star is sharp for Bob. For Alice (Bob) the fuzziness increases as the angular deviation from the yellow (green) star increases and the same stars are observed with different fuzziness because of the relative orientation of the $z$-axes of the two agents.}
    \label{fig:StarrySky}
\end{figure}

We now analyze how the situation changes in our novel noncommutative framework. We henceforth assume that Alice and Bob can choose the direction along which points can be sharp independently from one another. For definiteness, we focus on observer Alice first. 
In her task of mapping the starry sky, she focuses on a particular star, which she observes with one of her telescopes. This instrument is generally misaligned with respect to the $z$-axis she has chosen. In our framework, this means that there exists a state $\ket{\psi_A}$ connecting the relative orientation between the aforementioned telescope and the $z$-axis. Following our physical interpretation, $\ket{\psi_A}$ determines a quantum rotation matrix through the expectation value of $R_q$ and an uncertainty matrix defined by \eqref{incertezze}. This means that Alice cannot appreciate the relative orientation between the telescope and her $z$-axis with arbitrary precision since there will always be some intrinsic uncertainty given by the deformation of the $SU(2)$ group. As a consequence of this, she will not be able to sharply deduce the position of this star in the sky.
For that same star, this line of reasoning also applies to Bob but the state $\ket{\psi_B}$ will generally be different from $\ket{\psi_A}$, implying a different degree of fuzziness.
In turn, this procedure can be applied to any star in the sky so that Alice and Bob each have their own picture of the celestial sphere, as can be seen in \cref{fig:StarrySky}.
The striking result is that the uncertainty associated with each star is not an intrinsic property of spacetime but depends on the choices made by the observers. In particular, different choices of the $z$-axis give rise to different pictures of the starry sky. 
The subtle observation is that there is no way for Alice and Bob to define an objective celestial sphere, meaning that
the definition of space itself cannot be independent of the observer who reconstructs it. In this sense, we say that Alice and Bob are agents: we are abandoning the idea of objective space, replacing it with a notion of \emph{observed space} from which we cannot subtract the choices of the observer who infers it.

As mentioned earlier, it is important to note that, even though the $z$-axis is a preferred direction for observers (only its points can be sharp), our framework produces an isotropic description of space. Indeed, spatial rotations are a ($q$-deformed) symmetry.
There is no direction which is preferred {\it a priori}, and
the special role played by the z-axis in the reference frame of a given observer is only the result of the choices made by that observer in setting up their frame. The oberver chooses freely their preferred $z$-direction.
This should be contrasted with the case of standard spatial anisotropy, in which different directions have different properties {\it a priori}, independently of the choices made by observers, and invariance under spatial rotations  is lost.

\subsection{Numerical analysis}
To gain insight into these conceptual novelties and to better grasp the meaning of \cref{fig:StarrySky}, we now show some numerical examples in support of our claims. 
We start with Alices's point of view. Let her $z$-axis be aligned with a certain star ($\alpha$) and consider another star $(\beta)$ she wants to observe from a telescope, whose relative orientation with respect to the first star is described by the following state:
\begin{equation}
    \ket{\psi_A}=\ket{n=0;\phi=\frac{\pi}{2};\chi=0}\,.
\end{equation}
This state,already introduced in section \cref{semiclassical rotations} and which satisfies the ``classicality" conditions \eqref{criterio rotazione classica}, can be seen as a $q$-deformed rotation of $\pi$ around the $x$-axis.
The expectation value is obtained by substituting $\phi=\frac{\pi}{2}$ in  \eqref{axisplane} while the variances are the ones in \eqref{varxyplane} 
\begin{equation}\label{pi rotation}
    \expval{R_q}{\psi_A}=\begin{pmatrix}
    q & 0 & 0 \\
    0 & -q & 0 \\
    0 & 0 & -q^2
    \end{pmatrix} \quad \Delta R_q=\begin{pmatrix}
    \frac{1}{2}\sqrt{(1-q^2)(1-q^4)} & \frac{1}{2}\sqrt{(1-q^2)(1-q^4)} & \frac{(1+q^2)}{2}\sqrt{(1-q^2)} \\
    \frac{1}{2}\sqrt{(1-q^2)(1-q^4)} & \frac{1}{2}\sqrt{(1-q^2)(1-q^4)} & \frac{(1+q^2)}{2}\sqrt{(1-q^2)} \\
        \sqrt{q^2-q^4} & \sqrt{q^2-q^4} & 0
    \end{pmatrix}
\end{equation}
which at first order in $(1-q)$ give
\begin{equation}
\begin{gathered}
   \expval{R_q}{\psi_A}=\begin{pmatrix}
    1-(1-q) & 0 & 0 \\
    0 & -1+(1-q) & 0 \\
    0 & 0 & -1+2(1-q)
    \end{pmatrix}
    \\
    \Delta R_q=\begin{pmatrix}
   \sqrt{2}(1-q) & \sqrt{2}(1-q) & \sqrt{2(1-q)} \\
    \sqrt{2}(1-q) & \sqrt{2}(1-q) & \sqrt{2(1-q)} \\
        \sqrt{2(1-q)} & \sqrt{2(1-q)} & 0
    \end{pmatrix}
\end{gathered}
\end{equation}
Identifying Alice's $z$-axis with the vector $v=(0,0,1)$ and applying these matrices on $v$, the transformed vector $v'=(v'_1,v'_2,v'_3)$ will lie in the range 
\begin{equation}
    -\sqrt{2(1-q)}\leq v'_1\leq \sqrt{2(1-q)} \qquad -\sqrt{2(1-q)}\leq v'_2\leq \sqrt{2(1-q)} \qquad v'_3=-1+2(1-q)
\end{equation}
This quantitatively shows what we mean by fuzziness: the q-rotated vector $v'$ lies in a cone with aperture given by
\begin{equation}
\label{aperture}
    \Delta\alpha\approx 2\sqrt{2(1-q)}
\end{equation}
The agency feature of the model can be well understood if we now compare these results with the ones obtained if Alice chose to align her $z$-axis with star $\beta$. In this case she would have seen $\beta$ sharply and $\alpha$ under a cone with aperture \eqref{aperture}.
This line of reasoning can be extended when considering multiple stars: different states describing the relative orientation of these stars with respect to Alice's z-axis will produce different uncertainties in determining their direction. In turn, this is characterized by different cone apertures under which the stars are seen.
In \cref{table aperture} we exhibit some numerical examples expressing this feature. 

\begin{table}[h!]
\centering
\begin{tabular}{||c c||} 
 \hline
 $\pmb{\theta(n)}$ & \textbf{Aperture}\\ [0.5ex] 
 \hline\hline
  $0^\circ$ & $0$ \\ 
 \hline
 $\pm 45.275^\circ$ & $0.210$ \\ 
 \hline
 $\pm 90.560^\circ$ & $0.284$ \\
 \hline
 $\pm 137.518^\circ$ & $0.402$ \\
 \hline
 $\pm 180^\circ$ & $0.442$\\
 \hline
\end{tabular}
\caption{Dependence of the aperture on the observation angle $\theta(n)$ using $q=0.99$. The aperture is monotonically increasing as the angles grow in absolute value. The states used to obtain these results are constructed numerically in \cref{Numerical construction sec}. For consistency, the aperture for $180^\circ$ is not truncated at first order in $(1-q)$ as in \eqref{aperture}, but is computed numerically with \cref{pi rotation} with $q=0.99$.}
\label{table aperture}
\end{table}

From this analysis, it is worth noticing that the fuzziness grows as our quantized Euler angle increases, resulting in a starry sky inferred by Alice, similar to what is depicted in \cref{fig:StarrySky}. Of course, the analysis can also be repeated for Bob, who chooses a different z-axis, in principle, and observes the same stars as Alice. The states describing the relative orientation between these stars and his z-axis will be different from the ones characterizing Alice's frame. The degree of fuzziness he observes for a particular star will be different from the one assigned to it by Alice, resulting in a different inference of the starry sky, as shown in \cref{fig:StarrySky}.

\section{Conclusions}
In this work, we have given a physical interpretation of the $SU_q(2)$ quantum group and, from the properties of its representations, derived a framework in which one of the Euler angles becomes quantized and reference frames exhibit fuzziness properties. 

The nature of this fuzziness is distinct from the one usually associated with quantum references frames (QRFs) \cite{angelo2011physics,Giacomini:2017zju,Vanrietvelde:2018pgb,AhmadAli:2021ajb,delaHamette:2020dyi,delaHamette:2021oex, Castro-Ruiz:2019nnl, Hoehn:2019fsy,Lizzi:2018qaf}, which  arises describing the frame itself as an ordinary quantum system. The key difference is the source of the frame fuzziness: for QRFs it is governed by Planck's constant $\hbar$ and follows from  the standard laws of quantum mechanics, while in our scenario it is the deformation of spatial rotations, governed by the parameter $q$, that produces the fuzziness. Accordingly, the states encoding the fuzziness are associated with physical systems (including the frame) in the context of QRFs, while here they describe the relation between two frames. QRFs can be in relative superposition, however, the transformations between them act on the states of the physical systems directly, without being described by a quantum state of their own.

Classically, all observers can objectively reconstruct the sky. In our work, they are agents: using quantum rotations, they reconstruct their own version of the sky based on their choices of the $z$-axis and there is no operational procedure they can use to reconstruct it objectively. In fact, even if they were to exchange their data, they would need to $q$-rotate it in their coordinates, inevitably introducing the intrinsic uncertainty predicted by the $SU_q(2)$ model. 

The next step in this program would be to develop a fully relativistic and relational picture. To do so, we should first consider the case in which more observers are involved, understanding which states can describe the composition of deformed rotations. Then the framework should be extended to the translations and boosts sectors.

The full quantum regime should also be investigated. In our present work, we focused on a semi-classical regime, in which the expectation values and the variances of the $q$-deformed rotation matrix act on classical vectors, to derive some novel qualitative properties. This is a crucial step toward a description of a $q$-deformed version of the alignment procedure described in \cref{Protocol sec}.  To go further, we need a more developed characterization of physical objects like spinors/qubits and Stern-Gerlach apparata, as well as a consistent interpretational framework for such ``quantum mechanics on a quantum space(time)". In particular, we would need a description of qubits in a world in which rotations are described by $SU_q(2)$, and this would require facing several open questions: Is it consistent to just assume that the standard 1-qubit Hilbert space describes the spins that Alice sends to Bob? Or does consistency require generalizing the notion of spinors to some noncommutative objects (e.g., see \cite{Song:1991zz})?  What does the measurement procedure do to the state of the system? Is the relative orientation of Alice's and Bob's laboratories changed by their gaining information about each other while exchanging spins? Or is a quantum spacetime described by $SU_q(2)$ rotations a sufficiently mild deformation of classical spacetime that it still guarantees that two observers can study their relative orientation without changing it?

Besides their conceptual interest, investigating these possibilities
might open a path \cite{Amelino-Camelia:2008aez,Addazi:2021xuf} towards much-needed experimental tests, which in turn could guide further theoretical development of quantum gravity.

\section*{Acknowledgements}
We thank Francesco Minieri for the realization of the starry sky image in~\cref{fig:StarrySky}.
This work was made possible through support from the Foundational Questions Institute under grant number FQXi-RFP-1801A for the project ``Agency-dependent spacetime and spacetime-dependent agency'' (awarded to G.A.C.\ and P.H.). It has also received partial funding from Okinawa Institute of Science and Technology Graduate University.
F.M.\  has been partially supported by Agencia Estatal de Investigaci\'on (Spain)  under grant  PID2019-106802GB-I00/AEI/10.13039/501100011033.

\newpage
\appendix

\section{Classical $SU(2)$ parametrization and quantization ambiguity}\label{Quantization ambiguity sec}
The $SU(2)$ group is defined as the group of $2\cross 2$ unitary matrices with determinant equal to 1. A generic element  $U\in SU(2)$ can be parametrized as
\begin{equation}\label{su2classica}
    U = \begin{pmatrix}
    a & -c^*\\
    c & a^*
\end{pmatrix}\;\;,
\end{equation}
where $a$ and $c$ are two complex numbers satisfying
\begin{equation}
aa^* + cc^* =1 \;\;,
\end{equation}
constraining the degrees of freedom to three real numbers. A commonly used parametrization for $a$ and $c$ in terms of three real numbers is
\begin{equation}
    a = e^{i\eta}\cos{\frac{\theta}{2}} \qquad c = e^{i\delta} \sin{\frac{\theta}{2}}  \;\;,
    \label{rappresentazione classica a e c methods}
\end{equation}
with $\eta,\delta\in [0,2\pi)$ and $\theta\in [0,\pi)$. These three angles encode all the information needed to uniquely identify an  $SU(2)$ element and are useful in identifying the connection of $SU(2)$ to the rotation group $SO(3)$. Indeed, the canonical homomorphism between the two groups is realized via
\begin{equation}\label{MatriceR_AliceBob}
\tensor{R}{_{ij}} = \frac{1}{2}\Tr\qty{\sigma_i \, U\,\sigma_j\,U^\dagger} \equiv \frac{1}{2}\Tr\qty{U\,\sigma_j\,U^\dagger \, \sigma_i} \;\;,
\end{equation}
where $R_{ij}$ is a rotation matrix, $U$ is parametrized as in \eqref{su2classica} and $\sigma_i$ are the Hermitian Pauli matrices.
Writing \eqref{MatriceR_AliceBob} explicitly, we have
\begin{equation}
    R=\begin{pmatrix}
    \frac{1}{2}(a^2-c^2+(a^*)^2-(c^*)^2) & \frac{i}{2}(-a^2+c^2+(a^*)^2-(c^*)^2) & (a^*c+c^*a) \\
    \frac{i}{2}(a^2+c^2-(a^*)^2-(c^*)^2 )& \frac{1}{2}(a^2+c^2+(a^*)^2+(c^*)^2) & -i (a^*c-c^*a) \\
    -(ac+c^*a^*) & i(ac-c^*a^*) & 1-2cc^*
    \end{pmatrix} \;\; .
    \label{matrice rotazione classica}
\end{equation}
As anticipated in the results section, matrix \eqref{matrice rotazione quantum 1} indeed coincides with this classical one in the $q \rightarrow 1$ limit. 

Inserting parametrization \eqref{rappresentazione classica a e c methods} in \eqref{matrice rotazione classica} we obtain the rotation matrix in terms of trigonometric functions of the three real angles $\eta$, $\delta$ and $\theta$. These are simply a redefinition of the well known Euler angles $(\alpha,\beta,\gamma$)
\begin{equation}\label{euler angles}
    \theta=\beta \qquad \eta=\frac{\alpha+\gamma}{2} \qquad \delta=\frac{\pi}{2}-\frac{\alpha-\gamma}{2}\,.
\end{equation}
In terms of these, a generic rotation matrix is written as $R(\alpha,\beta,\gamma) = R_z(\alpha)R_x(\beta)R_z(\gamma)$, where $R_z$ and $R_x$ are rotations around the $z$-axis and $x$-axis respectively.

In the quantization procedure, we replace the classical $U$ with the deformed $U_q$ in \eqref{matrice rotazione classica}, thus replacing the complex numbers $a$ and $c$ with the operators satisfying \eqref{deformed group algebra}, and we leave the Pauli matrices untouched. It may be argued that a deformation of the Pauli matrices, such as the one studied in \cite{Song:1991zz}, should be considered here. However, doing so leads to non-hermitian operators for the entries of the resulting matrix. With our choice, not only do we get hermitian operators, but we reconstruct a matrix that matches the one obtained in \cite{Podles1995} upon doing a change of basis, and this guarantees that we obtain a representation of $SO_q(3)$ as explained shortly.

To discuss our results we used matrix \eqref{matrice rotazione quantum 1} which is obtained with the first ordering in \eqref{MatriceR_AliceBob}. Due to the noncommutative nature of operators $a$ and $c$, the cyclic property of the trace, expressed in \eqref{MatriceR_AliceBob} is not valid anymore. This leads to two possible definitions of the quantum rotation matrix. The alternative to \eqref{matrice rotazione quantum 1}, obtained using the second ordering in \eqref{MatriceR_AliceBob}, is

\begin{equation}
   P_q = \begin{pmatrix}
    \frac{1}{2}(a^2-qc^2+(a^*)^2-q(c^*)^2) & \frac{i}{2}(-a^2+qc^2+(a^*)^2-q(c^*)^2) &  q(a^*c+c^*a) \\
    \frac{i}{2}(a^2+qc^2-(a^*)^2-q(c^*)^2) & \frac{1}{2}(a^2+qc^2+(a^*)^2+q(c^*)^2) & - iq (a^*c-c^*a) \\
    -\frac{1+q^2}{2q}(ac+c^*a^*) & i\frac{1+q^2}{2q}(ac-c^*a^*) & 1-(1+q^2)cc^*
    \end{pmatrix}\,.
    \label{matrice rotazione quantum philipp}
\end{equation}
It turns out that $P_q$ and $R_q$ are actually similar matrices: $R_q=MP_qM^{-1}$, with

\begin{equation}
    M= \begin{pmatrix}
  2^{\frac{1}{3}}(\frac{q}{1+q^2})^{\frac{1}{3}} & 0 & 0\\
    0 & 2^{\frac{1}{3}}(\frac{q}{1+q^2})^{\frac{1}{3}} & 0 \\
   0 & 0 & \frac{(\frac{1}{q}+q)^{\frac{2}{3}}}{2^{\frac{2}{3}}}
    \end{pmatrix}\,.
    \label{matricecambio di base}
\end{equation}
This matrix (with its inverse) reduces to the identity at first order in $(1-q)$ so that the two matrices are equivalent from a phenomenological point of view. Indeed, since the quantum-gravity expectation is that $q$ is extremely close to 1, the search for experimental manifestations of our $q$-deformation is not likely to ever go beyond the leading $(1-q)$-order effects \cite{Amelino-Camelia:2008aez}. Notice that the two matrices come from an ambiguity of the ``quantization procedure", namely the ordering of the matrices in \eqref{MatriceR_AliceBob}. This is similar to what happens in standard quantum mechanics where different orderings in the classical observables give rise to different quantum operators which give the same classical limits.

Furthermore, the link with the quantum group $SO_q(3)$ is established as follows. The vector (co)-representation of $SU_q(2)$ is linked to the (co)-representation of the quantum group $SO_q(3)$ via the isomorphism $C(SO_q(3))\coloneqq C(SU_q(2)/\mathbb{Z}_2)$; the representation reads \cite{Podles1995}

\begin{equation}\label{math matrix}
d_1=
    \begin{pmatrix}
        (a^*)^2 & -(1+q^2)a^*c & -qc^2 \\
        c^*a^* & 1-(1+q^2)c^*c & ac \\
        -q(c^*)^2 & -(1+q^2)c^*a & a^2
    \end{pmatrix}\,.
\end{equation}
In the commutative limit, this matrix approaches a generic rotation matrix written in the so-called complex basis, and can be rewritten in the Cartesian basis by a similarity transformation, with  \begin{equation}
     N=\begin{pmatrix}
        -1 & -i & 0 \\
        0 & 0 & 1 \\
        -1 & i & 0
    \end{pmatrix}
\end{equation}
acting as the change of basis. By applying this transformation in the $q$-deformed case too, one can easily check that $N^{-1}d_1N$ is identically equal to \eqref{matrice rotazione quantum 1}. This clarifies the link to the $SO_q(3)$ quantum group. Matrices $P_q$ and $R_q$ are (co)-representations of $SO_q(3)$ since they can be obtained by a similarity transformation from \eqref{math matrix}. Therefore, our ansatz \eqref{trace} realizes the quantum version of the $SU(2)$ to $SO(3)$ homomorphism.

In closing this section, we offer some comments on another quantization ambiguity, concerning our definition of the angle $\theta(n)$ in \eqref{angoli quantum}. There we obtained the definition of $\sin\qty(\frac{\theta(n)}{2})$ by comparing the eigenvalues of $\pi(c)$ to the classical parametrization of $c$ in \eqref{rappresentazione classica a e c}. We find that this is consistent with the definition of $\cos\qty(\frac{\theta(n)}{2})$ in terms of $\pi(a^*a)$, which also acts diagonally. Indeed, on a generic eigenstate $\ket{n,\phi,\chi}$  (cf.\ \eqref{rappresentazione pi}) one has that
\begin{equation}
    \pi(a^*a)\ket{n,\phi,\chi}=(1-q^{2n})\ket{n,\phi,\chi}=\cos^2\qty(\frac{\theta(n)}{2})\ket{n,\phi,\chi}\,.
\end{equation}
Furthermore, this is in agreement with the fact that $a^*a+c^*c=\mathbbm{1}$, as can be inferred from \eqref{deformed group algebra}.
Nevertheless, it is noteworthy that since $a$ and $a^*$ do not commute (see \eqref{deformed group algebra}), one could obtain another result for $\theta(n)$ upon defining its cosine in terms of $\pi(aa^*)$, which also acts diagonally (cf.\ \eqref{rappresentazione pi}), rather than $\pi(a^*a)$. As one can easily check, this would lead to allowed values of $\theta$ which are the same as in our $\pi(a^*a)$ case, with the only difference that $\theta = \pi$ would be missing from the spectrum of $\theta(n)$. With this ordering we would have $\sin\qty(\frac{\theta(n)}{2})=q^{n+1}$ which is merely a shift of $n$ with respect to the ordering we adopted. Evidently, the main results and the main aspects of physical interpretation of our analysis are unaffected by this
quantization ambiguity.

\section{Numerical construction of semi-classical states}\label{Numerical construction sec}

In order to construct the semi-classical states, we have resorted to numerical computations, focusing on the case $q = 0.99$ as illustrative example. However, by increasing the value of $q$, we have checked that the below states satisfy the condition \eqref{criterio rotazione classica}, thus approximating classical rotations with increasing precision. 

We will focus on the $(y,z)$-plane setting $\chi=0$, $\phi=\frac{\pi}{2}$, as can be seen from \eqref{euler angles}. We want to construct states that semi-classically describe a deformed rotation of a certain angle in this plane. Since $n$ defines the only angle left, $\theta$, we choose to build superpositions \eqref{superposition} centered on particular values of $n$, dubbed $\bar{n}$, with coefficients $c_n$ multiplying the states $\ket{n,\phi,\chi}$ rapidly decreasing as $\theta(n)$ deviates  from $\theta\qty(\bar{n})$. Our ansatz is that these coefficients have the form of a discretized Gaussian distribution.

The variance is then chosen in the following way. Recalling that

\begin{equation}
    \theta(n)=2\arcsin{q^n}
\end{equation}
we can define $\Delta \Bar{n}$ as
\begin{equation}
    \Delta \Bar{n}\vcentcolon=\frac{dn}{d\theta}\Bigr|_{\theta \qty(\Bar{n})} \Delta \theta\,.
\end{equation}
We take $\Delta\theta=\frac{\pi}{2}-\arcsin{q}$, which is just half of the value of the maximum angular deviation and weigh it with the rate of change of $n$ with respect to $\theta$, approximated as the derivative. Therefore, the value of the variance depends on the central value $\Bar{n}$. This approximation becomes more and more accurate with increasing values of $n$ for which $\theta(n)$ becomes quasi-continuous.

We then define our superposition coefficients $c_n$ as

\begin{equation}\label{gaussian coefficients}
    c_n=\frac{e^{-\frac{(\Bar{n}-n)^2}{2\Delta \Bar{n}^2}}}{\sum_{n=0}^\infty e^{-\frac{(\Bar{n}-n)^2}{2\Delta \Bar{n}^2}}}\,.
\end{equation}
For computational reasons, we do not consider the full superposition going from $n=0$ to $n=\infty$ but we truncate the series by considering the 3-$\sigma$ range of our Gaussian. Namely, the sum goes from $n_{\text{min}}$ to $n_{\text{max}}$, where

\begin{equation}
    n_{\text{min}}-\Bar{n}=-3\Delta\Bar{n} \qquad n_{\text{max}}-\Bar{n}=3\Delta\Bar{n}\,.
\end{equation}
For relatively small values of $\Bar{n}$, the value for $n_{\text{min}}$ might be negative when considering the 3-$\sigma$ range. When this happens, we simply truncate the Gaussian and start the series from $n=0$. We have explicitly verified that this doesn't affect our results in a significant way. 

We have used this algorithm to construct the states used for the numerical analysis of \cref{table aperture}. In what follows, we show the results for the computation of the expectation values of the matrix elements $\tensor{\qty(R_q)}{_{ij}}$ with their relative uncertainties for states centered around $n=95,34,7,0$, corresponding to angles $\theta=45.275^\circ,90.560^\circ,137.518^\circ,\theta=180^\circ$.
\begin{itemize}
    \item $n=95$, $\theta=45.275^\circ$ \begin{equation}
\begin{gathered}
    \expval{R_q}{\psi(45.275^\circ)}=\begin{pmatrix}
    0.997 & 0.000 & 0.000 \\
    0.000 & 0.695 & 0.708 \\
    0.000 & -0.708 &0.698 
    \end{pmatrix}
    \\
    \Delta R_q=\begin{pmatrix}
 0.005 & 0.071 & 0.030 \\
    0.071 & 0.073 & 0.069 \\
        0.030 & 0.069 & 0.073
    \end{pmatrix}
\end{gathered}
\end{equation}

\item $n=34$, $\theta=90.560^\circ$ \begin{equation}
\begin{gathered}
   \expval{R_q}{\psi(90.560^\circ)}=\begin{pmatrix}
    0.990 & 0.000 & 0.000 \\
    0.000 & -0.015 & 0.990 \\
    0.000 & -0.990 &-0.005 
    \end{pmatrix}
    \\
    \Delta R_q=\begin{pmatrix}
 0.015 & 0.100 & 0.100 \\
    0.100 & 0.099 & 0.004 \\
        0.100 & 0.004 & 0.100
    \end{pmatrix}
\end{gathered}
\end{equation}

   \item $n=7$, $\theta=137.518^\circ$
\begin{equation}
\begin{gathered}
    \expval{R_q}{\psi(137.518^\circ)}=\begin{pmatrix}
    0.982 & 0.000 & 0.000 \\
    0.000 & -0.739 & 0.663 \\
    0.000 & -0.663 &-0.721 
    \end{pmatrix}
    \\
    \Delta R_q=\begin{pmatrix}
 0.025 & 0.067 & 0.173 \\
    0.067 & 0.067 & 0.074 \\
        0.173 & 0.074 & 0.067
    \end{pmatrix}
\end{gathered}
\end{equation}
\item n=0, $\theta=180^\circ$
\begin{equation}
\begin{gathered}
    \expval{R_q}{\psi(180^\circ)}=\begin{pmatrix}
    0.99 & 0 & 0 \\
    0 &- 0.99 & 0 \\
    0 & 0 &-0.99 
    \end{pmatrix}
    \\
    \Delta R_q=\begin{pmatrix}
 0.014 & 0.014 & 0.140 \\
   0.014 & 0.014 & 0.140 \\
        0.140 & 0.140 & 0
    \end{pmatrix}\,.
\end{gathered}
\end{equation}
\end{itemize}
To obtain the states describing $q$-deformations of rotations of negative angles $\zeta<0$, we just have to consider the same states but with $\phi = -\frac{\pi}{2}$. It is straightforward to show that for states of the form \eqref{superposition} with $\chi = 0$ having real superposition coefficients like \eqref{gaussian coefficients}, the map $\phi \mapsto - \phi$ exchanges $\tensor{\qty(R_q)}{_{23}}$ with $\tensor{\qty(R_q)}{_{32}}$, leaving all the other elements unchanged, and yields the same uncertainty matrix. Therefore the aperture of the cone in \cref{table aperture} is the same for opposite angles.

For the $\theta=0$ case, we use the state in the representation $\rho$ which gives the identity matrix in computing the expectation values which is simply given by $\ket{\chi=0} \in \mathcal{H}_{\rho}$. From \eqref{rotation z axis} we thus have
\begin{equation}
    \expval{R_q}{\psi(0^\circ)}= \mathbbm{1}_{3\times 3}
    \qquad \Delta R_q= \pmb{0}_{3\times3}\,.
\end{equation}

\section{Quantum protocol for the alignment of reference frames in  classical space}\label{Protocol sec}

Let us illustrate a simple (but not necessarily most efficient) protocol for two agents, who reside in a classical Euclidean spatial environment, but do not share a classical reference frame, to synchronize their spatial orientations by communicating qubits, i.e.\ $\rm{SU}(2)$ spinors. It is desirable to extend this protocol to the quantum group $\rm{SU}_q(2)$, however, we leave this for future research as it requires a better understanding of multi-qubit states and sequential rotations in the presence of the symmetry deformation.

\subsection{Qubits as $SU(2)$ spinors in standard quantum theory}

It is well-known that any single qubit state, represented as a density matrix $\rho$, can be expanded in the basis of the identity  $\mathbbm{1}$ and Hermitian Pauli matrices $\sigma_i$  as
\begin{equation}
\rho=\frac{1}{2}(\mathbbm{1}+\vec{r}\cdot\vec{\sigma})\,,\label{rhoexp}
\end{equation}
where  the expectation values of the spin observables give the components of the Bloch vector:
\begin{equation}
r_i = \langle\sigma_i\rangle=\Tr\qty{\rho\,\sigma_i}\,.\label{bloch}
\end{equation}
The latter relation holds since the Pauli matrices form an orthonormal basis
\begin{equation}
\Tr\qty{\sigma_\mu\,\sigma_\nu}=2\,\delta_{\mu\nu}\label{su2orth}\,,
\end{equation}
where $\sigma_0:=\mathbbm{1}$. As such, we can also define a 4-dimensional Bloch vector
\begin{equation}
r_\mu = \Tr\qty{\rho \,\sigma_\mu}
\end{equation}
whose zero-component
\begin{equation}
r_0=\Tr\rho=1\label{norm}
\end{equation}
is simply the normalization of the state.

Consider now an $SU(2)$ transformation of the quantum state. As it turns out, the density matrix \eqref{rhoexp} decomposes into an $SU(2)$-invariant singlet and an $SU(2)$-covariant triplet term, since for any matrix $U\in SU(2)$,
\begin{equation}\label{qubitTransfLaw}
\rho \to U\,\rho\,U^\dagger = \frac{1}{2}(\mathbf{1}+\vec{r}\cdot U\,\vec{\sigma}\,U^\dagger) = \frac{1}{2}(\mathbf{1}+\vec{r'}\cdot\vec{\sigma})\,,
\end{equation}
where now the transformed Bloch vector is
\begin{equation}
\vec{r'}=R\cdot\vec{r}
\end{equation}
with
\begin{equation}
\tensor{R}{_{ij}}=\frac{1}{2}\Tr \qty{U\sigma_j\,U^\dagger\,\sigma_i}\label{SO3}
\end{equation}
an element of $PSU(2)\simeq SO(3)$. In other words, if the quantum state is transformed by an $SU(2)$ matrix, the Bloch vector $\vec r$ that represents it transforms like a 3-dimensional vector under spatial rotations.

Clearly, the phase of the $SU(2)$ transformations is cancelled and the adjoint action of $SU(2)$ yields a spatial rotation in its fundamental representation. Hence, the normalization $r_0$ is an $SU(2)$-invariant, while $\vec{r}$ transforms in the adjoint representation of $SU(2)$ and thus under the fundamental representation of $SO(3)$.

In this manner, we can associate to every $SU(2)$ transformation of states or spin observables an $SO(3)$ spatial rotation. In particular, we can write the same qubit state relative to different bases of Pauli matrices $U\,\vec{\sigma}\,U^\dagger$ by simply rotating the corresponding Bloch vector accordingly,
\begin{equation}
\vec{r'}_\rho = \Tr\qty{\rho\,U\,\vec{\sigma}\,U^\dagger} = R^{-1}\cdot\vec{r}_\rho=R^{-1}\cdot\Tr\qty{\rho\,\vec{\sigma}}\,,\label{rotatesamebloch}
\end{equation}
where $\vec{r'}_\rho$ and $\vec{r}_\rho$ are both Bloch vectors of the \emph{same} quantum state $\rho$, however, written relative to different observable bases. Interpreting $U\,\vec{\sigma}\,U^\dagger$ and $\vec{\sigma}$ as being the observables corresponding to differently oriented sets of Stern-Gerlach devices in the lab, this rewriting allows us to express how observers with different spatially oriented reference frames will `see' the same quantum state of (possibly an ensemble of) a qubit.

\subsection{Alignment protocol between two laboratories by exchanging qubits}
\begin{figure}[t!]
    \centering
    \includegraphics[width=0.6\textwidth]{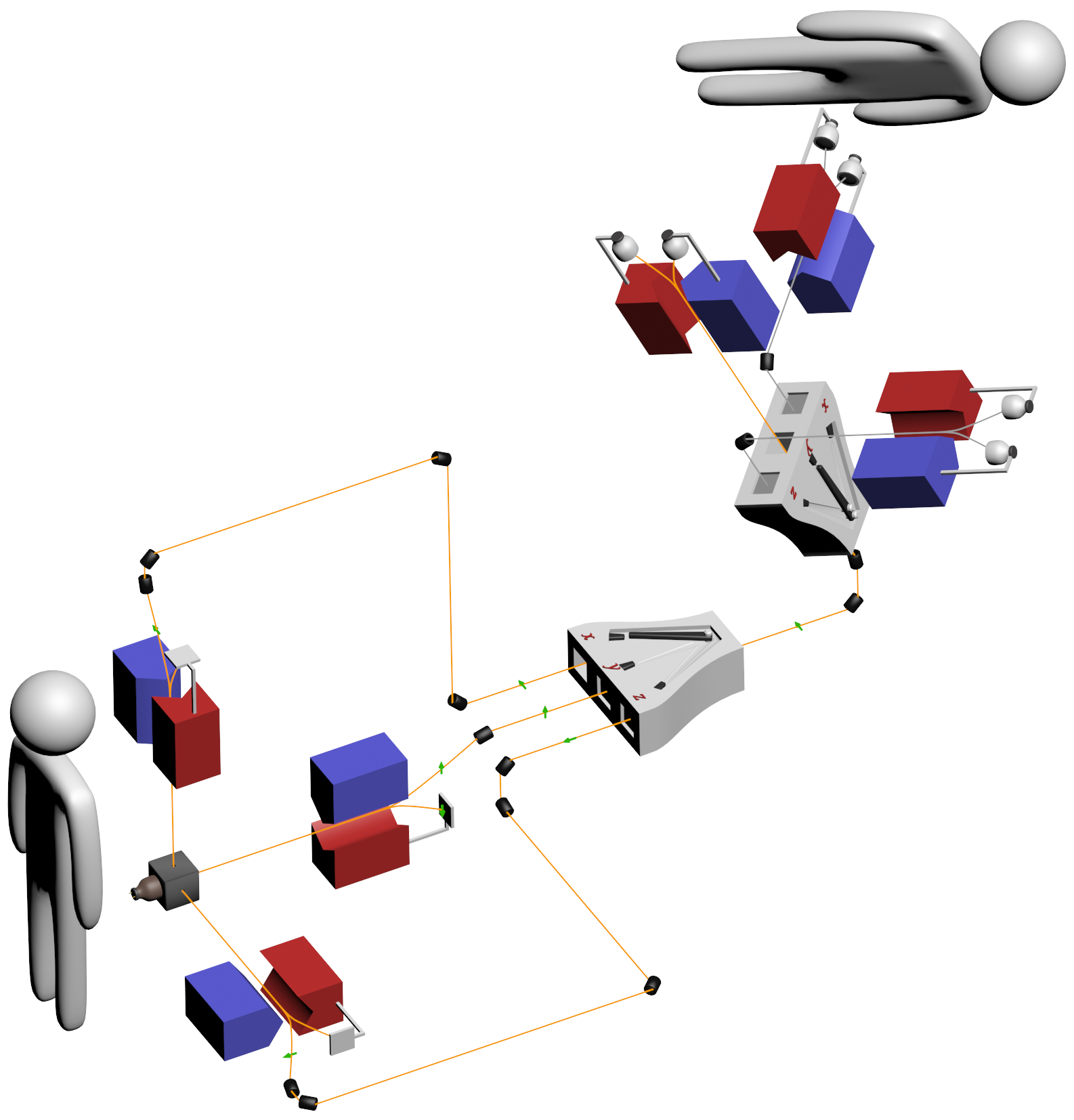}
    \caption{Alice (on the left) prepares a set of $N$ qubits (e.g., electron spins) in the spin-up eigenstate of her $x$ axis (e.g., by passing unpolarized electrons through a $x$-oriented Stern--Gerlach apparatus, and selecting only the ones that emerge with their spins up). She then sends these electrons to Bob, whose laboratory is rotated by an unknown amount with respect to hers. Bob divides these $N$ qubits in three groups, and sends each group through a machine that measures the spin along one of his three orthogonal axes (e.g., three perpendicular Stern--Gerlach apparata - in the picture, he is passing the electrons through a $y$-oriented machine). He then counts the number of spin-up and spin-down measurements that each machine reads, and calculates the expectation value of the corresponding observable. Repeating the experiment for a set of $N$ qubits that Alice selected to be polarized along the $y-$, and, respectively, $z-$axes allows Bob to build a statistics of the expectation values of the nine observables associated to each pair of choices of axes made by him and Alice. In the large-$N$ limit, these expectation values tend to the nine components of the rotation matrix $R$ that connects Alice's reference frame and Bob's. Notice that, in this illustrative picture the electron beams (in orange) are manipulated with some "mirrors" (the black cylinders) which are assumed not to have any effect on the qubit states.}
    \label{fig:my_label}
\end{figure}
Suppose that Alice and Bob have never met before and thus do not know how their reference frames are aligned with respect to each other. Supposing further that they can communicate classically, they can proceed as follows in order to figure out the relation between their descriptions. Alice prepares three ensembles of qubits, each prepared such that the ensemble state corresponds to a Bloch sphere basis vector. For example, she could prepare three ensembles of $N$ qubits each so that
\begin{equation}
E_{x_A} = \{\vec{r}_{x_A,n} = (1,0,0)\}_{n=1}^N\,,
\qquad
E_{y_A} = \{\vec{r}_{y_A,n} = (0,1,0)\}_{n=1}^N\,,
\qquad
E_{z_A} = \{\vec{r}_{z_A,n} = (0,0,1)\}_{n=1}^N\,,
\end{equation}
\emph{i.e.}, every qubit of the first ensemble is prepared to be in the `up in $x_A$-direction' pure state, where $x_A$ is Alice's $x$-direction, and similarly for the other two ensembles. Alice can then send Bob these three ensembles to perform state tomography on them. Bob could proceed as follows: he divides each ensemble in three subensembles and measures all qubits of the first subensemble with a Stern-Gerlach (SG) device oriented in $x_B$-direction, of the second in $y_B$-direction and of the third in $z_B$-direction. This will tell him how to write Alice's states relative to his basis and, via \eqref{rotatesamebloch},  what the relative $\rm{SO}(3)$ rotation between their spatial frames is. For example, relative to Alice, the ensemble state of $E_{x_A}$ appears as
\begin{equation}
\rho_{x_A} = \frac{1}{2}(\mathbbm{1}+\vec{r}_{x_A}\cdot\vec{\sigma}_A) = \frac{1}{2}(\mathbbm{1}+\sigma_{x_A})\,.
\end{equation}

Suppose now that Bob's measurement apparata are in the following relation with Alice's: $\vec{\sigma}_A = U\,\vec{\sigma}_B\,U^\dagger$.
Then, using \eqref{rotatesamebloch}, we can write the same ensemble state, but now relative to Bob as
\begin{equation}
\rho_{x_A}=\frac{1}{2}\qty(\mathbbm{1}+\vec{r}_{x_A}\cdot U\,\vec{\sigma_B}\,U^\dagger) = \frac{1}{2}\qty(\mathbbm{1}+\vec{r}_{x_B}\cdot\vec{\sigma}_B)\,,
\end{equation}
where
\begin{equation}
\vec{r}_{x_B} = R\cdot\vec{r}_{x_A}
\end{equation}
and
\begin{equation}\label{MatriceR_AliceBobappendix}
\tensor{R}{_{ij}} = \frac{1}{2}\Tr\qty{\qty(\sigma_A)_i \, U\,\qty(\sigma_A)_j\,U^\dagger} \equiv \frac{1}{2}\Tr\qty{U\,\qty(\sigma_A)_j\,U^\dagger \, \qty(\sigma_A)_i} \,,
\end{equation}
which is equivalent to \eqref{SO3} if we remember that $U$ are unitary matrices. Bob can measure $\vec{r}_{x_B} = \Tr\qty{\rho_{x_A}\vec{\sigma}_B}$ using tomography on $E_{x_A}$, which is why he has to split that ensemble into three subensembles to measure its qubits relative to a basis of SG devices.

Proceeding this ways also with $E_{y_A},E_{z_A}$, Bob can figure out the Bloch vector basis $\vec{r}_{x_B},\vec{r}_{y_B},\vec{r}_{z_B}$ of Alice's ensemble states relative to his frame and if Alice also tells him how she describes the same states, namely as $\vec{r}_{x_A},\vec{r}_{y_A},\vec{r}_{z_A}$, then they can simply compute $R$ from the relation of these two sets of bases. In principle, Bob has to perform an infinite number of measurements to determine the rotation matrix connecting his reference frame with Alice's with arbitrary precision. From this, the matrix \eqref{matrice rotazione classica} admits a direct, operational interpretation in terms of \eqref{MatriceR_AliceBobappendix}: it is the matrix of the asymptotic $N \to \infty$ values of the averages of the spin measurements that Bob performs on the qubit states sent to him by Alice.

\section{Representation $\rho$ as limit of representation $\pi$}\label{limit sec}

By adding a phase to the representation $\pi$ discussed in the mathematical literature \cite{Woronowicz87} we  have 

\begin{equation}
    \rho(a) \ket{\chi}=e^{i\chi}\ket{\chi}\qquad \rho(a^*)\ket{\chi}=e^{-i\chi}\ket{\chi} \qquad \rho(c)\ket{\chi}=\rho(c^*)\ket{\chi}=0
    \label{rappresentazione rho e rhostar suppl}
\end{equation}

\begin{eqnarray}
    \pi(a)\ket{n, \phi, \epsilon}&=&e^{i\epsilon}\cos{\qty(\frac{\theta(n)}{2})}\ket{n-1, \phi, \epsilon} \qquad \pi(c)\ket{n, \phi, \epsilon}=e^{i\phi}\sin{\qty(\frac{\theta(n)}{2})}\ket{n, \phi, \epsilon}
    \nonumber\\ \label{rappresentazione pi suppl}\\
    \pi(a^*)\ket{n, \phi, \epsilon}&=&e^{-i\epsilon}\sqrt{1-q^2\cos^2{\qty(\frac{\theta(n)}{2})}}\ket{n+1, \phi, \epsilon} \qquad \pi(c^*)\ket{n, \phi, \epsilon}=e^{-i\phi}\sin{\qty(\frac{\theta(n)}{2})}\ket{n, \phi, \epsilon}\nonumber
\end{eqnarray}
with our definition of quantized angle, $\theta(n)= 2 \arcsin{q^n}$. Looking at \eqref{rappresentazione rho e rhostar suppl} and \eqref{rappresentazione pi suppl}, it may seem that we have four Euler angles in the representations of $SU_q(2)$, namely the aforementioned $( \chi, \epsilon, \phi, \theta(n))$. This is a subtlety solved by observing that, with the additional $\epsilon$ phase that we introduced in the $\pi$-representation, the representation $\rho$ can be obtained as a somewhat trivial limit of representation $\pi$. Therefore, we have three physical degrees of freedom, with the angles $\chi$ and $\epsilon$ linked by this limit operation, so that we can ultimately set $\epsilon=\chi$.

To see this, we need to find a class of states such that
\begin{equation}
    \pi(a)\ket{\psi (\chi)} \rightarrow e^{i \chi} \ket{\psi (\chi)} \qquad , \qquad \pi(c)\ket{\psi (\chi)} \rightarrow 0
    \label{condizione}
\end{equation}
in this limit. We can show that states of the type (with the identification $\epsilon=\chi$)

\begin{equation}
    \ket{\psi_n} = \frac{1}{\sqrt{n+1}}\sum_{k=n}^{2n} \ket{k, \phi, \chi}
    \label{stato limite bordo}
\end{equation}
from the $\pi$-representation satisfy requirement \eqref{condizione} for $n\to\infty$.

In fact, for the operator $c$ we can immediately see that
\begin{equation}
    \pi(c)\ket{\psi_n} = e^{i \phi} \frac{1}{\sqrt{n+1}}\sum_{k=n}^{2n} q^{k}\ket{k, \phi, \chi}
\end{equation}
and therefore, since $q$ is slightly less than one 

\begin{equation}
    \expval{\pi(c^*)\pi(c)}{\psi_n} = \frac{1}{n+1}\frac{q^{2n}\left(1-q^{2n+2}\right)}{1-q^2} \rightarrow 0, \qquad n \rightarrow \infty.
\end{equation}
Hence, $\pi(c)\ket{\psi_n}\xrightarrow[]{n\to\infty}0$. The limit for the operator $a$ is more subtle; in this case we have that
\begin{equation}
    \pi(a)\ket{\psi_n} = \frac{e^{i \chi}}{\sqrt{n+1}} \sum_{k=n}^{2n} \sqrt{1-q^{2k}} \ket{k-1, \phi, \chi}
\end{equation}
and we consider the difference
\begin{equation}
   \ket{\psi'_n} = \pi(a)\ket{\psi_n} - e^{i \chi}\ket{\psi_n}\,.
   \label{stato differenza limite}
\end{equation}
To show that the limit $\pi(a)\ket{\psi_n} \rightarrow e^{i \chi} \ket{\psi}$ holds we will show that the succession of the norms of the states in \eqref{stato differenza limite} has $0$ limit. We have that
\begin{eqnarray}
    0 &\leq & \braket{\psi'_n} = \expval{\pi(a^*)\pi(a)}{\psi_n} + \braket{\psi_n} - 2 \Re{e^{i \chi}\expval{\pi(a^*)}{\psi_n}} = \nonumber \\
    &=& 1+ \frac{1-q^{2n}}{n+1} - \frac{1}{n+1} \sum_{k=n+1}^{2n} \left[2 \sqrt{1-q^{2k}} - \left(1-q^{2k}\right)\right] \leq  1+ \frac{1}{n+1} - \frac{1}{n+1} \sum_{k=n+1}^{2n} \left(1-q^{2k}\right) = \nonumber \\
    &=& \frac{1}{n+1}\left[2 - q^{2n+2}\frac{1-q^{2n}}{1-q^2}\right] \xrightarrow{n\rightarrow \infty} 0.
\end{eqnarray}

 This observation that representation $\pi$ converges for certain states to representation $\rho$ can be extended to generic functions of operators. In general, such functions can be written as a sum of powers of the operators $\pi(a)$, $\pi(a^*)$, $\pi(c)$ and $\pi(c^*)$. We now prove that these functions acting on a class of states in representation $\pi$ give, in a certain limit, the value that the same function yields in the limit of these sates with $\pi(a)$ replaced by $\rho(a)$ and so on.

In order to do so, we consider a slightly different class of states than the one in \eqref{stato limite bordo}. In particular we consider

\begin{equation}
    \ket{\Psi_n} = \frac{1}{\sqrt{n!+1}}\sum_{k=n!}^{2n!} \ket{k, \phi, \chi}\,.
    \label{stato limite bordo funzioni}
\end{equation}
The reason why we make this choice with the factorial will be clear in the following.

Consider first the case in which we have an operator of the form $\pi(c)^m \pi(a)^l$. If we act with this operator on the state \eqref{stato limite bordo funzioni} and we compute the norm of the state obtained in this way we get

\begin{equation}
\small
   0\leq \expval{\pi(a^*)^l\pi(c^*)^m \pi(c)^m\pi(a)^l}{\Psi_n} \leq \frac{q^{2m(n!-l)}\qty[1-q^{2m(n!+1)}]}{\qty(1-q^{2m})(n!+1)} \xrightarrow[m\neq 0]{n\rightarrow \infty} 0\,.
\end{equation}
The same argument holds if we have only powers of the operator $\pi(c)$ (i.e, $l=0$) or other combinations involving $\pi(a^*)$, $\pi(c^*)$ and different orderings. The case with $m=0$, in which the limit above is $1$, corresponds to a power of $\pi(a)$ which will be handled slightly differently in the following.

Now we turn to powers of the operator $\pi(a)$, i.e $\pi(a)^m$. We do something analogous to the argument used for $\pi(a)$, namely we define the difference state

\begin{equation}
    \ket{\Psi'_n} = \pi(a)^m\ket{\Psi_n}-e^{im\chi}\ket{\Psi_n}
\end{equation}
and show its norm goes to zero as $n \rightarrow \infty$. After some algebraic manipulations, this norm is found to satisfy

\begin{equation}
    0 \leq \braket{\Psi'_n} \leq \frac{2m}{n!+1}+\frac{1}{n!+1}\sum_{k=n!+m}^{2n!}\qty[q^{l(l+1)}q^{2kl}(-1)^l \qty(q^{-2k};q^2)_l-1] \overset{n\rightarrow \infty}{\longrightarrow} 0\,,
\end{equation}
where $\qty(q^{-2k};q^2)_l\vcentcolon=\prod_{s=0}^{l-1}\qty(1-q^{-2k+2s})$ is the $q$-Pochhammer symbol. The result of the limit comes from the fact that the term in the square brackets above is essentially just a sum of powers of $q$.

We have thus shown that a generic function of $\pi(a)$ and $\pi(c)$ operators applied to (a class of) states in the $\pi$ representation converges to the same function applied to a state in $\rho$ representation (a similar argument holds for functions of $\pi(a^*)$ and $\pi(c^*)$ operators). The reason why we considered a state of the form \eqref{stato limite bordo funzioni} is now clear: it is necessary to have a number of states in the superposition which goes to infinity faster than $m$ does in order to properly do the limit when considering generic functions, in which case we have a series in $m$.

The states \eqref{stato limite bordo} can be used also to show directly for the vectorial (co)-representation that it is possible to obtain the rotations about the $z$-axis as a limit in the representation $\pi$. In the same way, it is possible to obtain the identity, i.e.\ the null rotation, in the same representation.

To obtain a rotation about the $z$-axis, we must have $\phi = 0$ and we have to consider the large $n$ limit. In this limit, setting $\phi =0$, we can approximate our $\pi$ representation as
\begin{equation}
    \pi(a)\ket{n, \phi, \chi}=e^{i \chi}\ket{n-1, \phi, \chi} \qquad \pi(c)= \pi(c^*)=0 \qquad  \pi(a^*)\ket{n, \phi, \chi}= e^{-i \chi}\ket{n+1, \phi, \chi}\,.
\end{equation}
Therefore, using the states \eqref{stato limite bordo}, we obtain

\begin{equation}
    \expval{R_q}{\psi}=
    \begin{pmatrix}
       \frac{-1+n}{1+n}\cos{\chi} & \frac{-1+n}{1+n}\sin{\chi} & 0 \\
       -\frac{-1+n}{1+n}\sin{\chi} & \frac{-1+n}{1+n}\cos{\chi} & 0\\
       0 & 0 & 1
    \end{pmatrix}\,.
\label{limite matrice bordo}
\end{equation}
In the large $n$ limit, we see that \eqref{limite matrice bordo} approximates a rotation matrix about the $z$-axis with greater and greater precision and it can be shown that the variances are equal to $0$.
Moreover, $\chi=0$ in \eqref{limite matrice bordo} gives the identity matrix in the large $n$ limit.

\end{document}